# Peano High Impedance Surfaces

John McVay[1] , Ahmad Hoorfar[1] and Nader Engheta[2]

(1) Dept. of Electrical and Computer Engineering, Villanova University, Villanova, PA 19085

(2) Dept. of Electrical and Systems Engineering, University of Pennsylvania, Philadelphia, PA

19104

October 17, 2004

*Abstract*

Following our previous work on metamaterial high-impedance surfaces made of Hilbert curve inclusions, here we theoretically explore the performance of the high-impedance surfaces made of another form of space-filling curve known as the Peano curve. This metamaterial surface, formed by a 2-D periodic arrangement of Peano curve inclusions, is located above a conducting ground plane and is shown to exhibit a high surface impedance surface at certain specific frequencies. Our numerical study reveals the effect of the iteration order of the Peano curve, the surface height above the conducting ground plane and the separation distance between adjacent inclusions.

## 1. Introduction

In the mathematics literature, the Peano curve is considered a member of the family of curves known as the space-filling curves [Sagan, 1994]. It was introduced by Giuseppe Peano in 1890, and it was the first example of such a family of curves [Sagan, 1994]. The Peano curve, and in general, all space-filling curves, possess interesting features in that they can be made with an electrically long wire that is compacted in a small footprint. (See Fig. 1) In other words, as the iteration order of these curves increases, they maintain their footprint size while the lengths of these curves increase, implying that they can exhibit relatively long resonant wavelengths with respect to the linear dimension of their footprints.

Artificial magnetic conducting surfaces have high surface impedances, which result in a reflection coefficient of $\Gamma \simeq +1$, when illuminated with a plane wave, instead of the typical $\Gamma \simeq -1$ for a conventional perfectly electric conducting (PEC) surface (see e.g., [Sievenpiper et al., 1999; Yang and Rahmat-Samii, 2002, Zhang et al., 2002; Gonzalo et al., 1999; Hansen, 2002; Engheta, 2002]). These structures have various applications in the antenna design [Sievenpiper et al., 1999; Yang and Rahmat-Samii, 2002, Zhang et al., 2002; Gonzalo et al., 1999; Hansen, 2002] and in thin absorbing screens [Engheta, 2002].

In the present work, we numerically analyze the performance of a metamaterial surface in which many Peano-curve elements in a 2-D periodic arrangement are placed above a ground plane. In our previous work, we have investigated the behavior of the analogous Hilbert surface as a high-impedance ground plane [McVay et al., 1993; McVay et al., 2004].  Here we extend our work to the case of Peano surfaces.

## 2. Scattering from a Single Peano Curve Inclusion.

To understand the electromagnetic properties of the Peano space-filling curves, single Peano curve elements of varying order were studied in free-space, under the influence of a normally incident plane wave, using the Method of Moments. Each curve was assumed to have a footprint of 30mm x 30mm, and  modeled as a thin metallic wire of radius 0.125 mm Each structure was excited with a normally incident plane wave with the electric field in the plane containing the curve. The frequency of the plane wave was varied and the maximum value of the current induced on each element was evaluated as a function of the excitation frequency. Two different plane-wave polarizations were studied, where the electric field was polarized in the x-direction, Ex, and the y-direction, Ey. In Figures 2 and 3, the maximum value of the current magnitude induced on each of the Peano Curve elements is shown versus frequency for both the Ex and Ey polarizations, respectively. Also shown in these figures are the current distributions at the resonant frequency for each case. It can be seen from Fig. 2 that for the Ex polarization case, a first resonance, which corresponds to the lower frequency resonances for the Ey polarizations, becomes much less dominant as the iteration order of the curve is increased and, as it will be shown later that this resonance vanishes completely in reflection from surfaces made of the higher order Peano inclusions.

In Fig. 4 the 30 mm side dimension of each Peano curve, when it is normalized with respect to the resonant wavelength λres, is plotted as a function of iteration order of the curve. The corresponding bandwidth, which is defined here by the difference in frequency values where the maximum values of the current magnitude falls to 0.707 times the maximum of maxima, is shown here. (This frequency difference is normalized with respect to the resonant frequency.). (Note that for the x-polarized cases, only the "dominant" resonance indicated in Fig. 2 by the normalized value of 1, is utilized to evaluate the data in Fig. 4). It can be seen that as the order of the curve is increased, the electrical footprint of the curve decreases since the resonant frequency decreases, as expected and evident from Figs. 2 and 3. The dependence of the resonant frequency on the polarization of the incident wave can also be seen from this figure. For the cases where the incident electric field is polarized in the y-direction, the resonant frequency is approximately 1/3 that of the x-polarized case. Further insight into this polarization dependence can be obtained from the current distributions along the Peano curve, shown in Figs. 2 and 3.

From the results presented above, it can be seen that the Peano-curve element can resonate at frequencies where the footprint of the curve can be considered electrically very small. The higher the order of the curve, the lower the resonant frequency and thus, the smaller the footprint of the curve with respect to the resonant wavelength. The cost of achieving such a compact resonant structure is clearly seen in the effect on the bandwidth. This effect on bandwidth is in general expected, as a resonant structure becomes effectively smaller with respect to the resonant wavelength. Such effect was also observed in design of electrically small antennas patterned after Peano or Hilbert curve elements [Zhu et al., 2003] as well as in the scattering from Hilbert curve inclusions [McVay et al., 2004]. As compared to a Hilbert curve element, however, a Peano curve element of identical footprint and iteration order resonates at a much lower frequency, albeit at the expense of a smaller bandwidth, due to the higher compression rate of the Peano curve algorithm.

### 3. High-Impedance Surfaces made of Peano Curve Inclusions.

To construct a surface of Peano curve inclusions, the Peano curve elements can be placed in a planer, two-dimensional array as shown in Fig. 5. To evaluate the scattering properties of this array that is infinitely extent in its plane, a periodic Method of Moments code was utilized. In this case, each element was modeled as a thin metallic strip with a strip width of 0.5 mm. The footprint dimensions remain identical to the previous cases (30 x 30 mm). The Peano array was placed a short distance (15 mm) above a conducting ground-plane of infinite extent. Again, a time-harmonic, normally incident plane wave was utilized to excite this structure, and the reflection coefficient from the surface was numerically evaluated as a function of frequency. Different polarizations were again used.

Fig. 6 shows the magnitude and phase of the reflection coefficient, versus frequency, for the Peano surface comprised of Peano curves of order 2 located at a height of 15 mm above the conducting ground-plane and a separation distance of 3.75 mm between inclusions within the array. The structure was illuminated with a normally incident plane-wave, polarized in the x and y directions separately. Since a ground-plane of infinite extent is present under the Peano surface, the magnitude of the reflection coefficient is always unity since all energy is always reflected. It should be noted here that all metals present in our numerical simulations are considered lossless therefore conductor losses are not taken into account. Also, the substrate between the Peano surface and the conducting ground plane is assumed to be air and thus no dielectric losses are present.

In Fig. 6a, the phase of the reflection coefficient at 0.5 GHz is shown to be approximately 180 degrees. As the frequency increases, this phase passes through 0 degrees and goes towards 180 degrees. At the frequency where the phase is 0 degrees (1.53 GHz), the Peano surface above the ground plane achieves an overall reflection coefficient of +1 and therefore acts as a high-impedance surface (HIS) (i.e. artificial magnetic conductor). Far away from this resonance denoted by $F_{HIS}$, this surface has an overall reflection coefficient of -1, and thus acts as a traditional electric conducting ground plane. It can be noted here that the footprints of the inclusions are approximately 0.063 and 0.153 $\lambda_{HIS}$ at the respective $F_{HIS}$ and the height above the ground-plane is approximately 0.031 and 0.076 $\lambda_{HIS}$ and thus both the inclusions and the height above the substrate are considered to be electrically small at resonance, for both polarizations.

We have also performed similar analyses for the surfaces made of the Peano curve inclusions of orders 1 and 3. The corresponding resonant frequencies and corresponding bandwidths are shown in Fig. 7. The bandwidths here are defined by the frequency values where the reflection coefficient phase falls between ± 90 degrees.

### *4. Effects of substrate height and inter-element spacing.*

In order to investigate these effects the height of a Peano surface composed of Peano curves of order 2 with a inter-element separation distance of 3.75 mm, was varied from 5 mm to 15 mm in steps of 1 mm. The maximum height was chosen such that the surface can still be considered to be electrically close to the ground plane for the smallest operating wavelength. Fig. 8 shows the $F_{HIS}$ frequency as well as the ±90° bandwidth as a function of the height of the surface above the conducting ground-plane. It can be seen that for the x-polarized cases, the resonant frequency decreases and the bandwidth increases as the height above the ground-plane increases. Very little change is found for the y-polarized resonances, which is due to the fact that since these resonances occur at lower frequencies, the relative change in height with respect to the resonant wavelength is less pronounced.

A parametric study was also performed with respect to the separation distance (inter-element spacing) between the Peano curves inclusions within the infinite two dimensional array. The separation distances were varied from 1 mm to 15 mm in steps of 2 mm. The results of this study are shown in Fig. 9, which shows the $F_{HIS}$ frequency as well as the ±90° bandwidth as a function of the separation distance. We can see that for the x-polarized cases, the resonant frequency increases whereas the bandwidth decreases as the separation distance increases. This trend is also present for the y-polarized resonances albeit less pronounced due to the fact that these resonances occur at lower frequencies, and again the relative change in the separation distance with respect to the resonant wavelength, is less noticeable.

## 5. Conclusions.

In this work, using numerical methods we have explored the reflection properties of a normally incident wave from a surface made of a 2-D periodic arrangement of Peano curve inclusions above a conducting ground plane. We have shown that this surface can act as a high impedance surface within a certain frequency band, as shown in Fig. 6. The frequency at which this surface becomes a high-impedance surface is mainly related to the iteration order of the Peano element. This frequency and the associated bandwidth are influenced by the height of the surface above the ground plane and to the lesser extent by the distance of separation between adjacent Peano elements within the array. We are investigating the role of the Peano surfaces in the antenna applications and we will report the results in due time.

### Acknowledgements

This work was supported in parts by the Office of Naval Research under Grant N000140410619 and by DARPA under Grant MDA972-02-1-0022. The content of this paper does not necessarily reflect the position or the policy of the U.S. Government, and no official endorsement should be inferred.

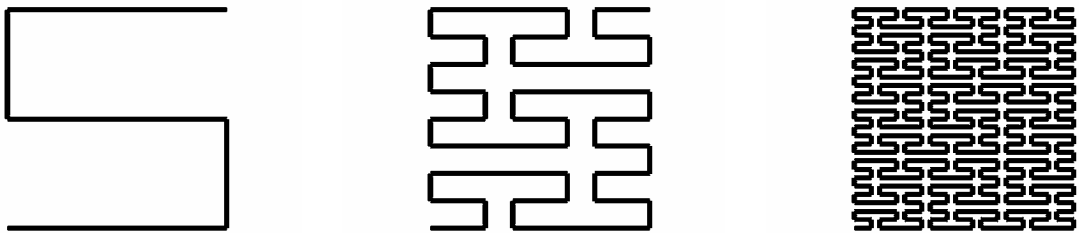

Figure 1: Peano Space-Filling Curves, Orders 1, 2 and 3

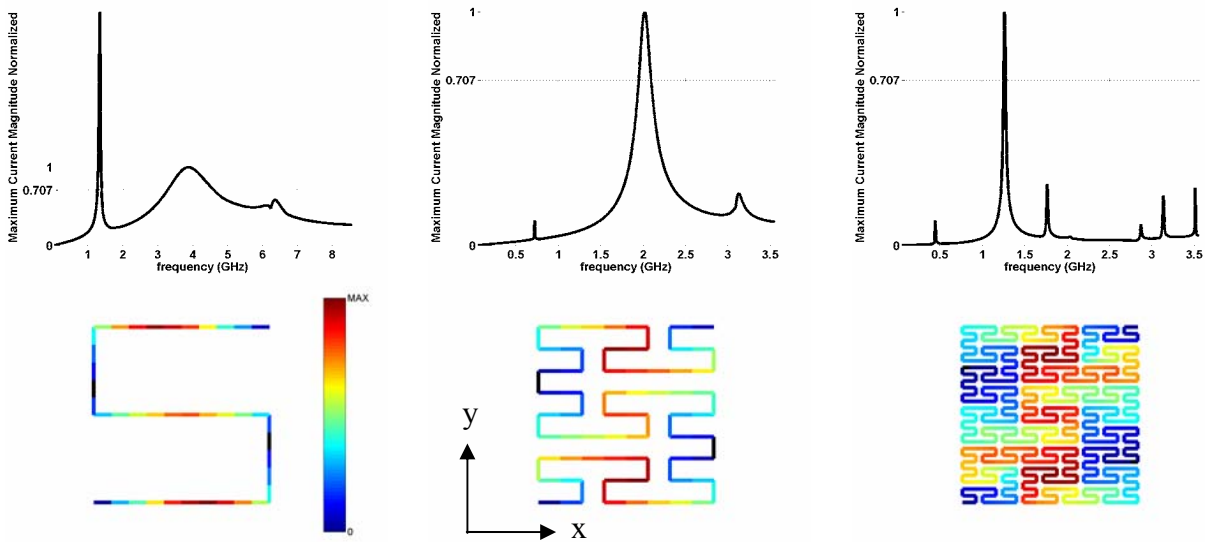

Figure 2: (Top row) Distribution of the maximum values of magnitude of the induced current versus frequency, (Bottom row) Distribution of the magnitude of induced current along the Peano curve at the resonant frequency; for Peano Curves of orders 1, 2 and 3, with Ex-polarized normally incident plane wave

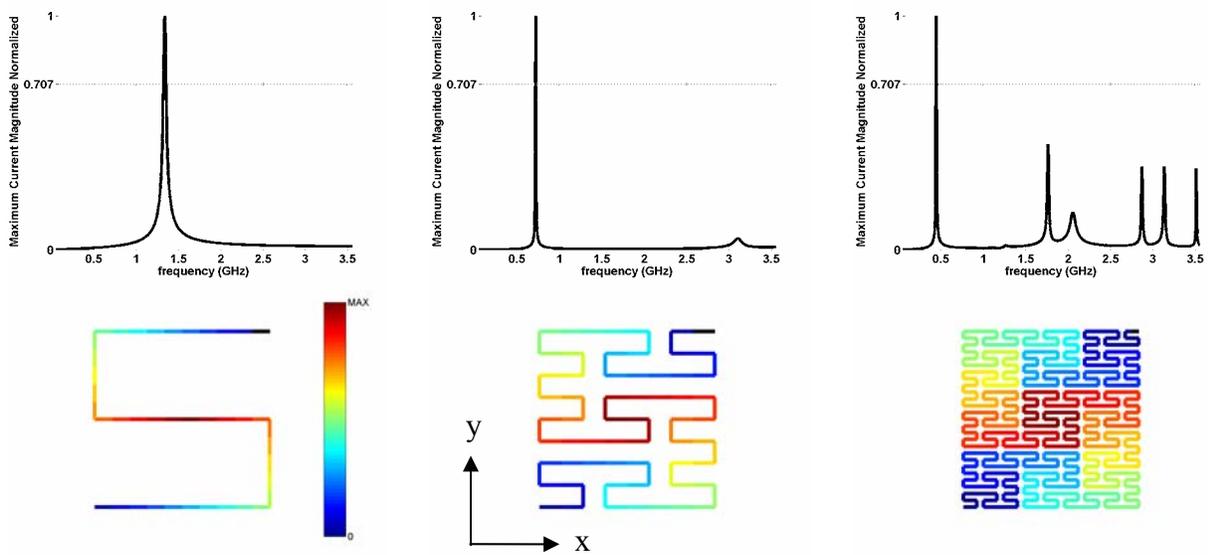

Figure 3: Similar to the caption of Fig. 2, but for Ey polarization

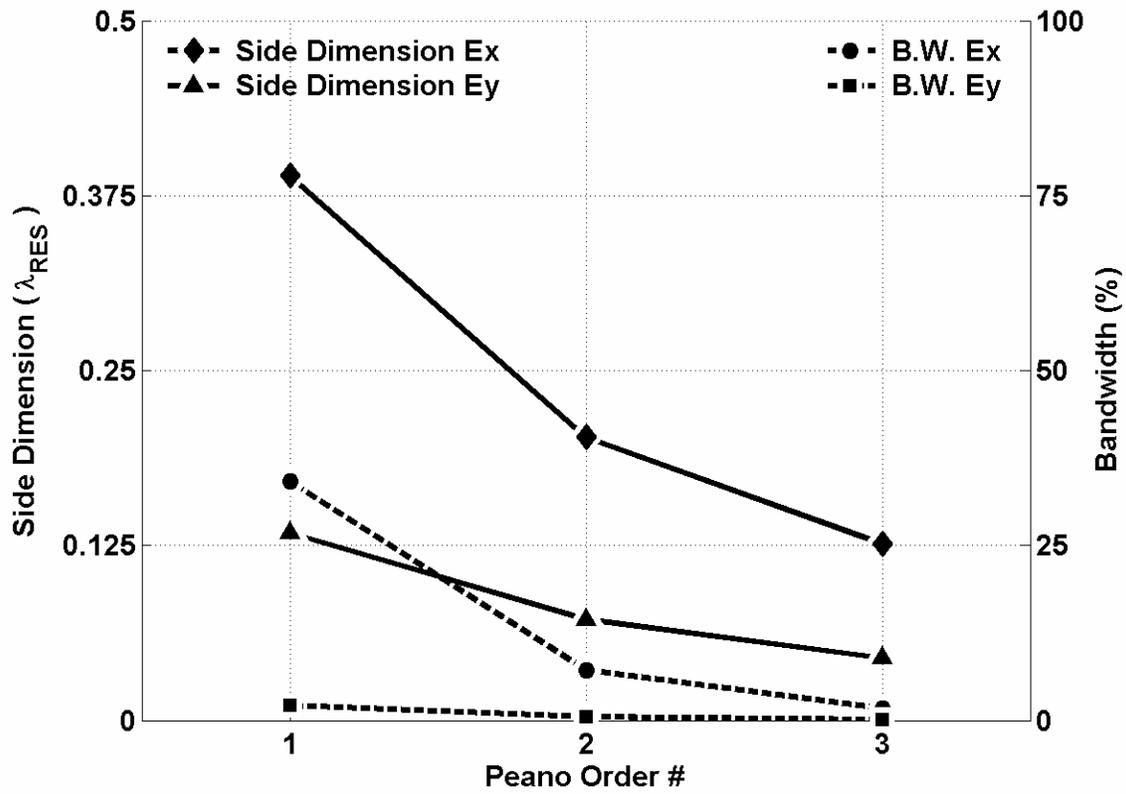

Figure 4: Side Dimensions (with respect to the resonant wavelength λres)  and relative
Bandwidths of maximum value of induced current on the Peano curve

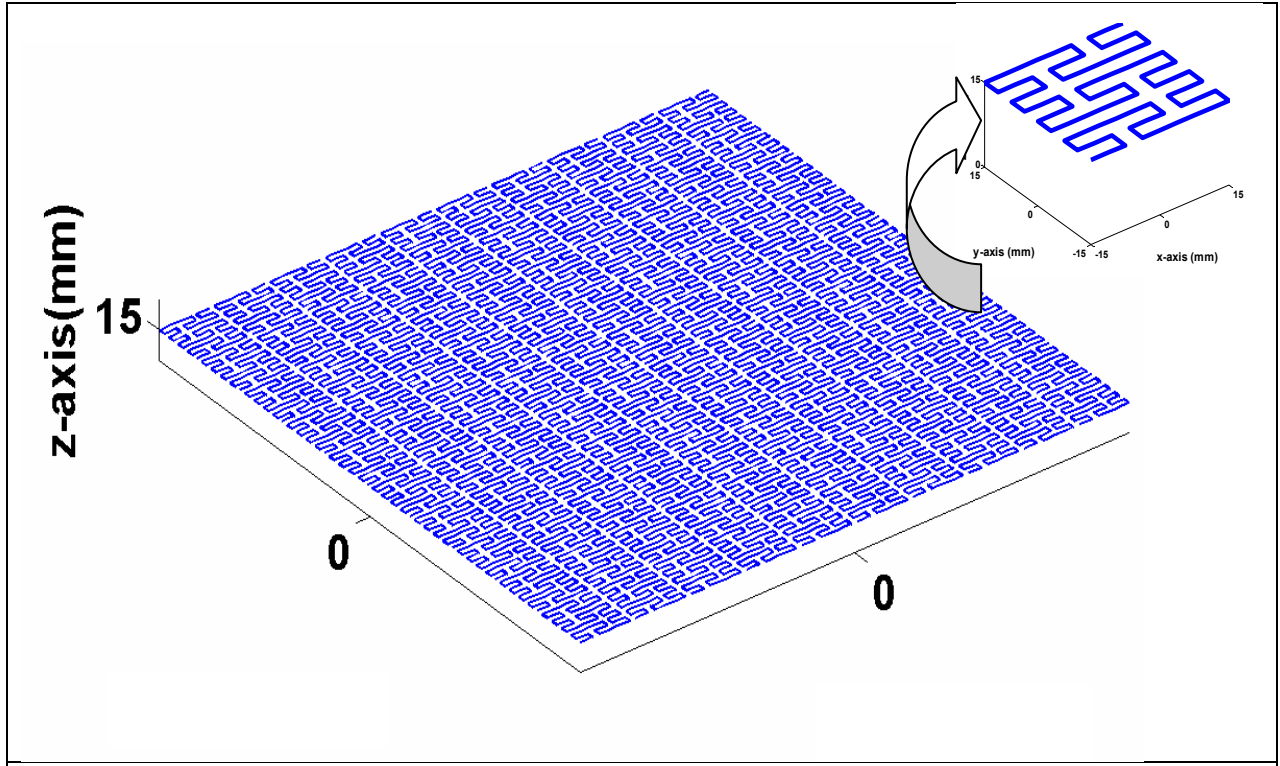

Figure 5: 3-Dimensional view of the Peano Surface of Order 2 above a conducting ground plane.

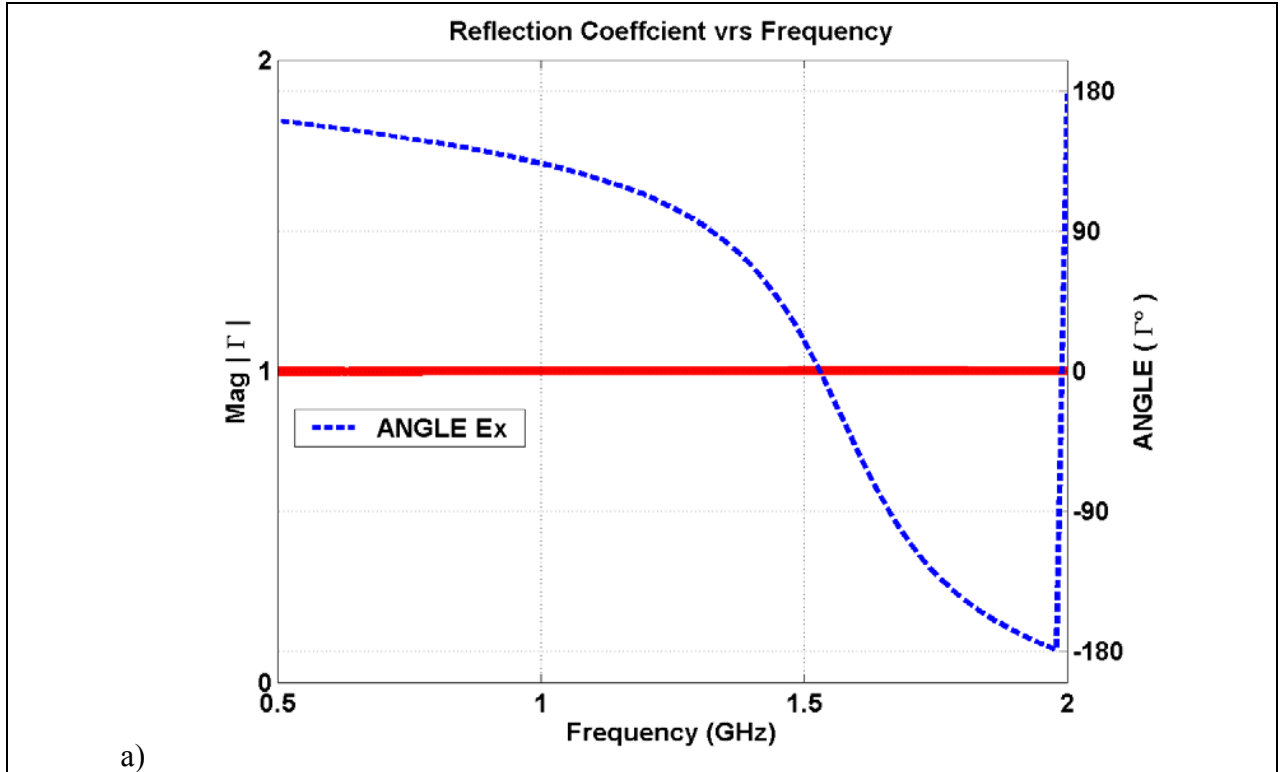

a)

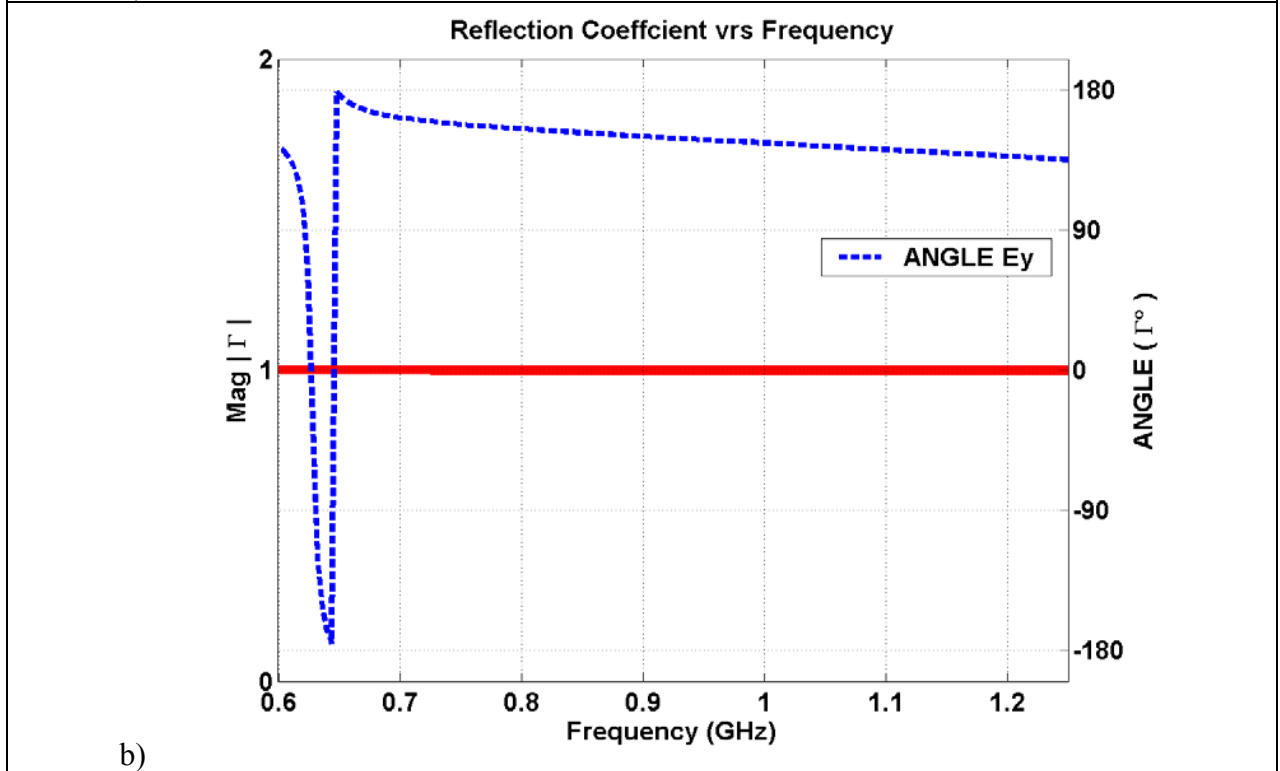

b)

Figure 6: Magnitude and phase of the Reflection Coefficient from a Peano surface of order 2 above a conducting ground plane, for the normally incident wave with polarizations in the x (a) and y (b) directions.

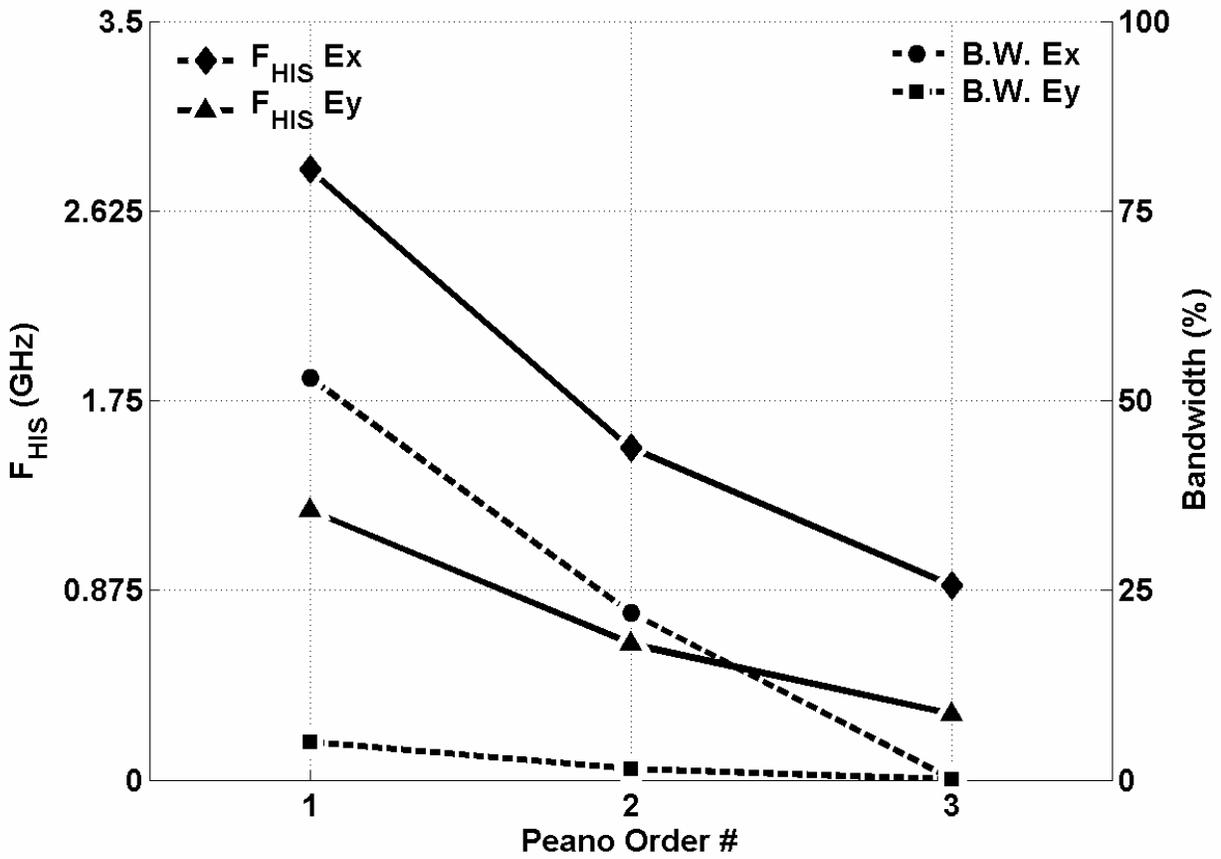

Figure 7: $F_{HIS}$ and Bandwidths of the Peano surface above the conducting ground plane, for iteration orders 1-3

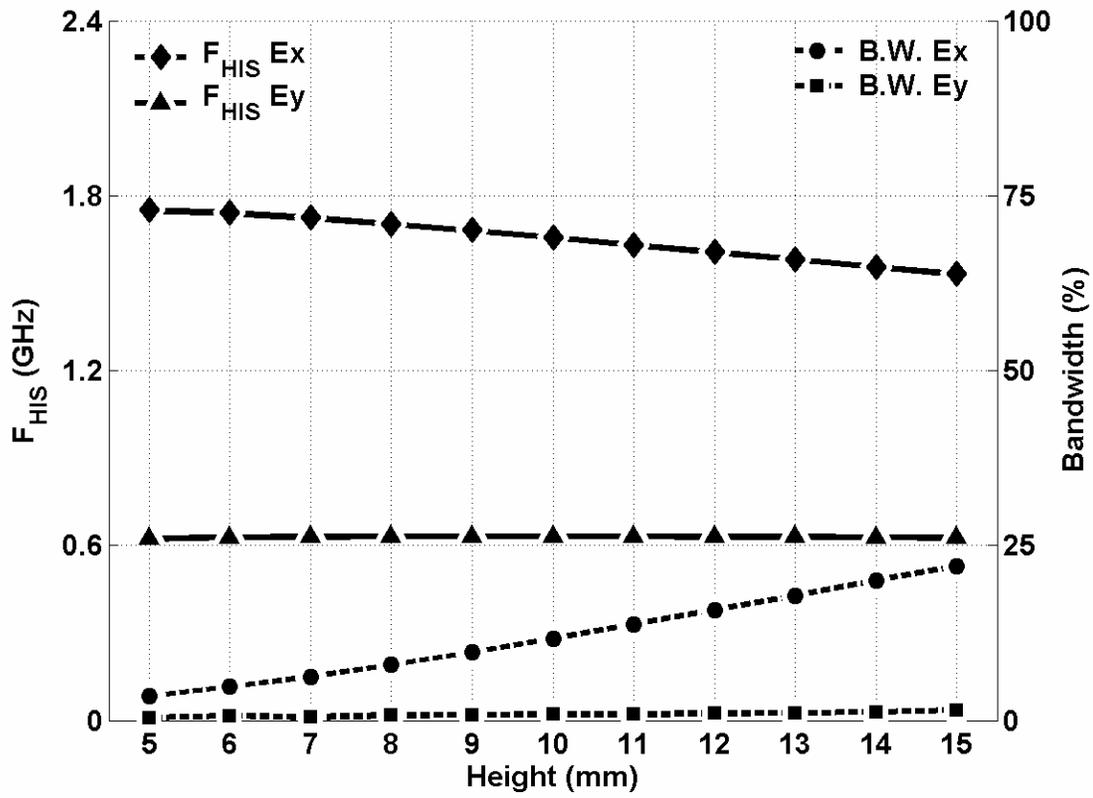

Figure 8: $F_{HIS}$ and Bandwidths versus Height for Peano surface of Order 2

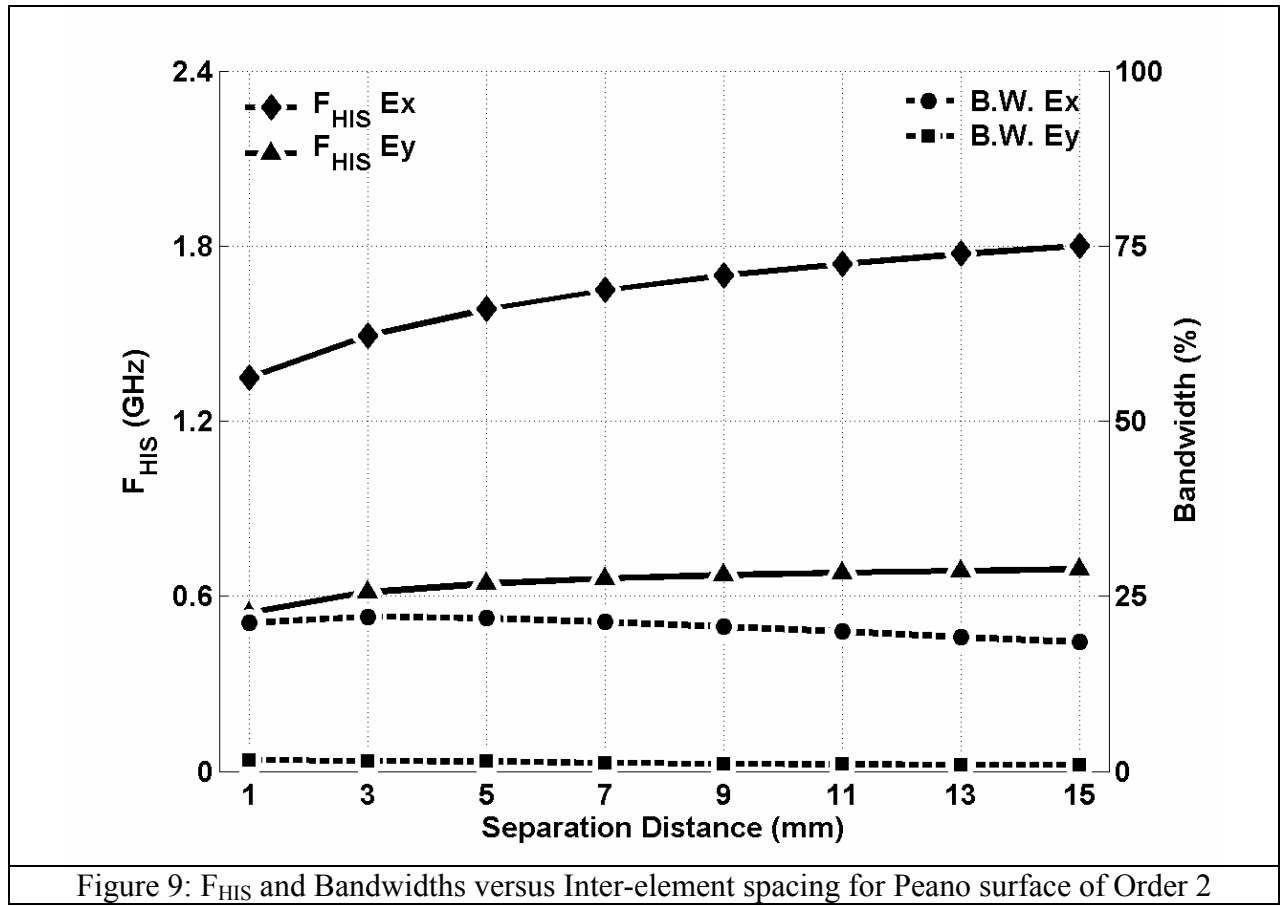

Figure 9: $F_{HIS}$ and Bandwidths versus Inter-element spacing for Peano surface of Order 2